\documentclass[12pt,a4paper]{article}
\usepackage{amsmath}
\usepackage{latexsym}
\makeatletter
\def\rddots{\mathinner{\mkern1mu\raise\p@%
    \vbox{\kern7\p@\hbox{.}}\mkern2mu%
    \raise4\p@\hbox{.}\mkern2mu\raise7\p@\hbox{.}\mkern1mu}}
\makeatother
\newcommand{\ket}[1]{{\vert{#1}\rangle}}
\newcommand{\bra}[1]{{\langle{#1}\vert}}

\newcommand{\fukuso}{{\mathbf C}}

\begin{document}

\title{\sl An Approximate Solution of the Jaynes--Cummings Model 
with Dissipation II : Another Approach}
\author{
  Kazuyuki FUJII
  \thanks{E-mail address : fujii@yokohama-cu.ac.jp }\quad and\ \ 
  Tatsuo SUZUKI
  \thanks{E-mail address : suzukita@sic.shibaura-it.ac.jp }\\
  ${}^{*}$Department of Mathematical Sciences\\
  Yokohama City University\\
  Yokohama, 236--0027\\
  Japan\\
  ${}^{\dagger}$Department of Mathematical Sciences\\
  College of Systems Engineering and Science\\
  Shibaura Institute of Technology\\
  Saitama, 337--8570\\
  Japan\\
  }
\date{}
\maketitle
\begin{abstract}
  In the preceding paper (arXiv:1103.0329 [math-ph]) we treated 
  the Jaynes--Cummings model with dissipation and gave an approximate 
  solution to the master equation for the density operator under {the 
  general setting} by making use of the Zassenhaus expansion.
  
  However, to obtain a compact form of the approximate solution (which 
  is in general complicated infinite series) is very hard when an initial 
  condition is given. To overcome this difficulty we develop 
  another approach and obtain a compact approximate solution 
  when some initial condition is given.
\end{abstract}
%

%
%
%
%
This paper is a sequel to the preceding one \cite{FS}. 
In the paper we treat the Jaynes--Cummings model with dissipation 
(or the quantum damped Jaynes--Cummings model in our terminology) 
once more and study the structure of general solution 
from a mathematical point of view. 

We want to apply it to Quantum Computation and Quantum Control 
which are our final target \cite{FHKW1}, \cite{FHKW2}. 
As a general introduction to these topics see for example \cite{BP} and 
\cite{WS}. 

We expect that our study will become a starting point to study more 
sophisticated models with dissipation in a near future. 

Let us start with the following phenomenological master equation 
for the density operator of the atom--cavity system 
in \cite{Scala et al-1} :
\begin{equation}
\label{eq:Q-D Jaynes-Cummings}
\frac{\partial}{\partial t}\rho=-i[H,\rho]
+
{\mu}
\left\{a\rho a^{\dagger}-\frac{1}{2}(a^{\dagger}a\rho+\rho a^{\dagger}a)\right\}
+
{\nu}
\left\{a^{\dagger}\rho{a}-\frac{1}{2}(aa^{\dagger}\rho+\rho aa^{\dagger})\right\}
\end{equation}
where $H$ (for simplicity we write $H$ not $H_{JC}$ in \cite{FS}) is 
the well--known Jaynes-Cummings Hamiltonian (see \cite{JC}) given by
\begin{eqnarray}
\label{eq:Jaynes-Cummings}
H
&=&
\frac{\omega_{0}}{2}\sigma_{3}\otimes {\bf 1}+ 
\omega_{0}1_{2}\otimes a^{\dagger}a +
\Omega\left(\sigma_{+}\otimes a+\sigma_{-}\otimes a^{\dagger}\right)
\nonumber \\
&=&
\left(
  \begin{array}{cc}
    \frac{\omega_{0}}{2}+\omega_{0}N & \Omega a             \\
    \Omega a^{\dagger} & -\frac{\omega_{0}}{2}+\omega_{0}N
  \end{array}
\right)
\end{eqnarray}
with
\[
\sigma_{+} = 
\left(
  \begin{array}{cc}
    0 & 1 \\
    0 & 0
  \end{array}
\right), \quad 
\sigma_{-} = 
\left(
  \begin{array}{cc}
    0 & 0 \\
    1 & 0
  \end{array}
\right), \quad 
\sigma_{3} = 
\left(
  \begin{array}{cc}
    1 & 0  \\
    0 & -1
  \end{array}
\right), \quad 
1_{2} = 
\left(
  \begin{array}{cc}
    1 & 0 \\
    0 & 1
  \end{array}
\right)
\]
, and $a$ and $a^{\dagger}$ are the annihilation and creation 
operators of an electro--magnetic field mode in a cavity, 
$N\equiv a^{\dagger}a$ is the number operator, and 
$\mu$ and $\nu$ ($\mu > \nu \geq 0$) are some constants 
depending on it (for example, a damping rate of the cavity mode). 

Note that the density operator $\rho$ is in $M(2;\fukuso)\otimes M({\cal F})=
M(2;M({\cal F}))$, namely
\begin{equation}
\label{eq:density operator}
\rho=
\left(
  \begin{array}{cc}
    \rho_{00} & \rho_{01}  \\
    \rho_{10} & \rho_{11}
  \end{array}
\right)
\in M(2;M({\cal F}))
\end{equation}
where $M({\cal F})$ is the set of all operators on the Fock space 
${\cal F}$ defined by
\begin{eqnarray*}
{\cal F}
&\equiv &\mbox{Vect}_{\fukuso}\{\ket{0},\ket{1},\ket{2},\ket{3},
\cdots \} \\
&=&
\left\{\sum_{n=0}^{\infty}c_{n}\ket{n}\ |\ \sum_{n=0}^{\infty}|c_{n}|^{2}<\infty\right\}
;\quad \ket{n}=\frac{(a^{\dagger})^{n}}{\sqrt{n!}}\ket{0}
\end{eqnarray*}
and ${\bf 1}$ in (\ref{eq:Jaynes-Cummings}) is the identity operator.

Now, we decompose (\ref{eq:Jaynes-Cummings}) into diagonal and 
off--diagonal parts
\begin{equation}
\label{eq:decomposition}
H
=H_{d}+H_{off}
=\left(
  \begin{array}{cc}
    \frac{\omega_{0}}{2}+\omega_{0}N &                                                \\
                                                 & -\frac{\omega_{0}}{2}+\omega_{0}N
  \end{array}
\right)
+
\left(
  \begin{array}{cc}
                                & \Omega a \\
    \Omega a^{\dagger} & 
  \end{array}
\right)
\end{equation}
and rewrite (\ref{eq:Q-D Jaynes-Cummings}) as
\begin{equation}
\label{eq:Jaynes-Cummings decomposed}
\frac{\partial}{\partial t}\rho=-i[H_{d},\rho]
+
{\mu}
\left\{a\rho a^{\dagger}-\frac{1}{2}(a^{\dagger}a\rho+\rho a^{\dagger}a)\right\}
+
{\nu}
\left\{a^{\dagger}\rho{a}-\frac{1}{2}(aa^{\dagger}\rho+\rho aa^{\dagger})\right\}
-i[H_{off},\rho].
\end{equation}
Namely, the main part is 
\[
-i[H_{d},\rho]
+
{\mu}
\left\{a\rho a^{\dagger}-\frac{1}{2}(a^{\dagger}a\rho+\rho a^{\dagger}a)\right\}
+
{\nu}
\left\{a^{\dagger}\rho{a}-\frac{1}{2}(aa^{\dagger}\rho+\rho aa^{\dagger})\right\}
\]
, while a kind of perturbed one is
\[
-i[H_{off},\rho].
\]
This is the main difference between \cite{FS} and this paper. 
Although the form may be not standard it is very useful to calculate  
when some initial conditions are given.

Let us write down the equation (\ref{eq:Jaynes-Cummings decomposed}) 
in a component-wise manner. Then
\begin{footnotesize}
\begin{eqnarray}
\label{eq:Each Component}
\dot{\rho}_{00}
&=&
-i\omega_{0}(N\rho_{00}-\rho_{00}N)+
{\mu}\left\{a\rho_{00}a^{\dagger}-
\frac{1}{2}(a^{\dagger}a\rho_{00}+\rho_{00}a^{\dagger}a)\right\}
+
{\nu}\left\{a^{\dagger}\rho_{00}{a}-
\frac{1}{2}(aa^{\dagger}\rho_{00}+\rho_{00}aa^{\dagger})\right\}
\nonumber \\
&&
-i\Omega(a\rho_{10}-\rho_{01}a^{\dagger}), 
\nonumber \\
\dot{\rho}_{01}
&=&
-i\omega_{0}(\rho_{01}+N\rho_{01}-\rho_{01}N)+
{\mu}\left\{a\rho_{01}a^{\dagger}-
\frac{1}{2}(a^{\dagger}a\rho_{01}+\rho_{01}a^{\dagger}a)\right\}
+
{\nu}\left\{a^{\dagger}\rho_{01}{a}-
\frac{1}{2}(aa^{\dagger}\rho_{01}+\rho_{01}aa^{\dagger})\right\}
\nonumber \\
&&
-i\Omega(a\rho_{11}-\rho_{00}a),
\nonumber \\
\dot{\rho}_{10}
&=&
-i\omega_{0}(-\rho_{10}+N\rho_{10}-\rho_{10}N)+
{\mu}\left\{a\rho_{01}a^{\dagger}-
\frac{1}{2}(a^{\dagger}a\rho_{01}+\rho_{01}a^{\dagger}a)\right\}
+
{\nu}\left\{a^{\dagger}\rho_{01}{a}-
\frac{1}{2}(aa^{\dagger}\rho_{01}+\rho_{01}aa^{\dagger})\right\}
\nonumber \\
&&
-i\Omega(a^{\dagger}\rho_{00}-\rho_{11}a^{\dagger}),
\nonumber \\
\dot{\rho}_{11}
&=&
-i\omega_{0}(N\rho_{11}-\rho_{11}N)+
{\mu}\left\{a\rho_{11}a^{\dagger}-
\frac{1}{2}(a^{\dagger}a\rho_{11}+\rho_{11}a^{\dagger}a)\right\}
+
{\nu}\left\{a^{\dagger}\rho_{11}{a}-
\frac{1}{2}(aa^{\dagger}\rho_{11}+\rho_{11}aa^{\dagger})\right\}
\nonumber \\
&&
-i\Omega(a^{\dagger}\rho_{01}-\rho_{10}a)
\end{eqnarray}
\end{footnotesize}
where $\dot{\rho}_{ij}=(\partial/\partial t)\rho_{ij}$ as usual.

Here we use some technique used in \cite{FS} (see also 
\cite{EFS} and \cite{FS2}), which is very useful in some case. 
For a matrix $X=(x_{ij})\in M({\cal F})$ 
\[X=
\left(
\begin{array}{cccc}
x_{00} & x_{01} & x_{02} & \cdots  \\
x_{10} & x_{11} & x_{12} & \cdots  \\
x_{20} & x_{21} & x_{22} & \cdots  \\
\vdots & \vdots & \vdots & \ddots
\end{array}
\right)
\]
we correspond to the vector $\widehat{X}\in 
{\cal F}^{\mbox{dim}_{\fukuso}{\cal F}}$ as
\begin{equation}
\label{eq:correspondence}
X=(x_{ij})\ \longrightarrow\ 
\widehat{X}=(x_{00},x_{01},x_{02},\cdots;x_{10},x_{11},x_{12},\cdots;
x_{20},x_{21},x_{22},\cdots;\cdots \cdots)^{T}
\end{equation}
where $T$ means the transpose. Then the following formula
\begin{equation}
\label{eq:well--known formula}
\widehat{EXF}=(E\otimes F^{T})\widehat{X}
\end{equation}
holds for $E,F,X\in M({\cal F})$.

This and equations (\ref{eq:Each Component}) give
\begin{eqnarray}
\label{eq:as a Vector}
\dot{\hat{\rho}}_{00}
&=&
\left[
-i\omega_{0}
\left(N\otimes {\bf 1}-{\bf 1}\otimes N\right)+
{\mu}
\left\{
a\otimes (a^{\dagger})^{T}-
\frac{1}{2}(a^{\dagger}a\otimes {\bf 1}+{\bf 1}\otimes a^{\dagger}a)
\right\}+
\right. \nonumber \\
&&
\left. \ 
{\nu}
\left\{a^{\dagger}\otimes a^{T}-
\frac{1}{2}(aa^{\dagger}\otimes {\bf 1}+{\bf 1}\otimes aa^{\dagger})
\right\}
\right]
\hat{\rho}_{00}
-i\Omega
\left( 
a\otimes {\bf 1}\hat{\rho}_{10}-{\bf 1}\otimes (a^{\dagger})^{T}\hat{\rho}_{01}
\right),
\nonumber \\
\dot{\hat{\rho}}_{01}
&=&
\left[
-i\omega_{0}
\left({\bf 1}\otimes {\bf 1}+N\otimes {\bf 1}-{\bf 1}\otimes N\right)+  
{\mu}
\left\{
a\otimes (a^{\dagger})^{T}-
\frac{1}{2}(a^{\dagger}a\otimes {\bf 1}+{\bf 1}\otimes a^{\dagger}a)
\right\}+
\right. \nonumber \\
&&
\left.\ 
{\nu}
\left\{a^{\dagger}\otimes a^{T}-
\frac{1}{2}(aa^{\dagger}\otimes {\bf 1}+{\bf 1}\otimes aa^{\dagger})
\right\}
\right]
\hat{\rho}_{01}
-i\Omega
\left( 
a\otimes {\bf 1}\hat{\rho}_{11}-{\bf 1}\otimes a^{T}\hat{\rho}_{00}
\right),
\nonumber \\
\dot{\hat{\rho}}_{10}
&=&
\left[
-i\omega_{0}
\left(-{\bf 1}\otimes {\bf 1}+N\otimes {\bf 1}-{\bf 1}\otimes N\right)+
{\mu}
\left\{
a\otimes (a^{\dagger})^{T}-
\frac{1}{2}(a^{\dagger}a\otimes {\bf 1}+{\bf 1}\otimes a^{\dagger}a)
\right\}+
\right.\nonumber \\
&&
\left.\ 
{\nu}
\left\{a^{\dagger}\otimes a^{T}-
\frac{1}{2}(aa^{\dagger}\otimes {\bf 1}+{\bf 1}\otimes aa^{\dagger})
\right\}
\right]
\hat{\rho}_{10}
-i\Omega
\left( 
a^{\dagger}\otimes {\bf 1}\hat{\rho}_{00}-{\bf 1}\otimes (a^{\dagger})^{T}\hat{\rho}_{11}
\right),
\nonumber \\
\dot{\hat{\rho}}_{11}
&=&
\left[
-i\omega_{0}
\left(N\otimes {\bf 1}-{\bf 1}\otimes N\right)+
{\mu}
\left\{
a\otimes (a^{\dagger})^{T}-
\frac{1}{2}(a^{\dagger}a\otimes {\bf 1}+{\bf 1}\otimes a^{\dagger}a)
\right\}+
\right. \nonumber \\
&&
\left.\ 
{\nu}
\left\{a^{\dagger}\otimes a^{T}-
\frac{1}{2}(aa^{\dagger}\otimes {\bf 1}+{\bf 1}\otimes aa^{\dagger})
\right\}
\right]
\hat{\rho}_{11}
-i\Omega
\left( 
a^{\dagger}\otimes {\bf 1}\hat{\rho}_{01}-{\bf 1}\otimes a^{T}\hat{\rho}_{10}
\right)
\end{eqnarray}
because ${\bf 1}$ and $N=a^{\dagger}a$ are diagonal (${\bf 1}^{T}=
{\bf 1},\ N^{T}=N$).

Next, in order to rewrite matrix elements in terms of Lie algebraic 
notations used in \cite{EFS} we set
\begin{eqnarray}
K_{+}&=&a^{\dagger}\otimes a^{T},\ \ 
K_{-}=a\otimes (a^{\dagger})^{T},\ \ 
K_{3}=\frac{1}{2}(N\otimes {\bf 1}+{\bf 1}\otimes N+{\bf 1}\otimes {\bf 1}), 
\nonumber \\
K_{0}&=&N\otimes {\bf 1}-{\bf 1}\otimes N.
\end{eqnarray}
Then it is easy to see
\begin{eqnarray}
\label{eq:relations I}
&&(K_{+})^{\dagger}=K_{-},\ \ (K_{3})^{\dagger}=K_{3}, 
\ \ (K_{0})^{\dagger}=K_{0}, \nonumber \\
&&[K_{3},K_{+}]=K_{+},\ \ 
[K_{3},K_{-}]=-K_{-},\ \ 
[K_{+},K_{-}]=-2K_{3}, \nonumber \\
&&[K_{0},K_{+}]=[K_{0},K_{-}]=[K_{0},K_{3}]=0.
\end{eqnarray}
Namely, $\{K_{+},K_{-},K_{3}\}$ are generators of the Lie algebra 
$su(1,1)$, see for example \cite{KF1} as a general introduction.

If we set from (\ref{eq:density operator}) 
\begin{equation}
\rho=
\left(
  \begin{array}{cc}
    \rho_{00} & \rho_{01}  \\
    \rho_{10} & \rho_{11}
  \end{array}
\right)
\ \Longrightarrow\ 
\hat{\rho}=
\left(
\begin{array}{c}
\hat{\rho}_{00} \\
\hat{\rho}_{01} \\
\hat{\rho}_{10} \\
\hat{\rho}_{11} 
\end{array}
\right)
\end{equation}
we obtain the following ``canonical" form
\begin{equation}
\label{eq:final form}
\frac{\partial}{\partial t}\hat{\rho}=(X+Y)\hat{\rho}
\end{equation}
with
\begin{eqnarray*}
\label{eq:}
&&X=
\left(
\begin{array}{cccc}
-i\omega_{0}K_{0}+L & 0 & 0 & 0                  \\
0 & -i\omega_{0}-i\omega_{0}K_{0}+L & 0 & 0 \\
0 & 0 & i\omega_{0}-i\omega_{0}K_{0}+L & 0   \\
0 & 0 & 0 & -i\omega_{0}K_{0}+L
\end{array}
\right),  \\
&&Y=
-i\Omega\left(
\begin{array}{cccc}
0 & -{\bf 1}\otimes (a^{\dagger})^{T} & a\otimes {\bf 1} & 0 \\
-{\bf 1}\otimes a^{T} & 0 & 0 & a\otimes {\bf 1} \\
a^{\dagger}\otimes {\bf 1} & 0 & 0 & -{\bf 1}\otimes (a^{\dagger})^{T} \\
0 & a^{\dagger}\otimes {\bf 1} & -{\bf 1}\otimes a^{T} & 0
\end{array}
\right) 
\end{eqnarray*}
and
\begin{eqnarray*}
L
&=&
{\mu}
\left\{
a\otimes (a^{\dagger})^{T}-
\frac{1}{2}(a^{\dagger}a\otimes {\bf 1}+{\bf 1}\otimes a^{\dagger}a)
\right\}+
{\nu}
\left\{a^{\dagger}\otimes a^{T}-
\frac{1}{2}(aa^{\dagger}\otimes {\bf 1}+{\bf 1}\otimes aa^{\dagger})
\right\}  \\
&=&
\frac{\mu-\nu}{2}+\nu K_{+}+\mu K_{-}-(\mu+\nu)K_{3}
\end{eqnarray*}
where the fundamental relation $aa^{\dagger}=a^{\dagger}a+{\bf 1}=N+{\bf 1}$ 
and a simplified notation $\omega_{0}$ in place of 
$\omega_{0}{\bf 1}\otimes {\bf 1}$ have been used.

\noindent
Let us note once more that $X$ is not anti--hermitian, while 
$Y$ is anti--hermitian.

Since the general solution of the equation (\ref{eq:final form}) is given by
\begin{equation}
\label{eq:general solution}
\hat{\rho}(t)=e^{t(X+Y)}\hat{\rho}(0)
\end{equation}
in a formal way we must calculate the term\ $e^{t(X+Y)}$, 
which is in general not easy (to obtain a compact form is almost 
impossible). 
For that the following Zassenhaus formula is convenient.

\vspace{3mm}\noindent
{\bf Zassenhaus Formula}\ \ We have an expansion
\begin{equation}
\label{eq:Zassenhaus formula}
e^{t(A+B)}=
\cdots 
e^{-\frac{t^{3}}{6}\{2[[A,B],B]+[[A,B],A]\}}
e^{\frac{t^{2}}{2}[A,B]}
e^{tB}
e^{tA}.
\end{equation}
The formula is a bit different from that of \cite{CZ}. 

In this paper we use the approximation
\begin{equation}
\label{eq:approximate formula}
e^{t(X+Y)}\approx e^{\frac{t^{2}}{2}[X,Y]}e^{tY}e^{tX}.
\end{equation}

\vspace{3mm}\noindent
Let us calculate each term explicitly.

\vspace{5mm}\noindent
[I]\ First, we calculate $e^{tX}$. The result is
\begin{equation}
\label{eq:}
e^{tX}
=
\left(
\begin{array}{cccc}
e^{t(-i\omega_{0}K_{0}+L)} & 0 & 0 & 0                       \\
0 & e^{-i\omega_{0}t}e^{t(-i\omega_{0}K_{0}+L)} & 0 & 0 \\
0 & 0 & e^{i\omega_{0}t}e^{t(-i\omega_{0}K_{0}+L)} & 0   \\
0 & 0 & 0 & e^{t(-i\omega_{0}K_{0}+L)}
\end{array}
\right)
\end{equation}
and fortunately in \cite{EFS} the term $e^{t(-i\omega_{0}K_{0}+L)}$ 
has been calculated exactly. Namely,
\begin{equation}
\label{eq:disentangling formula}
e^{t(-i\omega_{0}K_{0}+L)}
=
e^{\frac{\mu-\nu}{2}t}
e^{G(t)K_{+}}e^{-i\omega_{0}tK_{0}-2\log(F(t))K_{3}}e^{E(t)K_{-}}
\end{equation}
where
\begin{eqnarray}
\label{eq:fundamental functions}
E(t)&=&\frac{\frac{2\mu}{\mu-\nu}\sinh\left(\frac{\mu-\nu}{2}t\right)}
     {\cosh\left(\frac{\mu-\nu}{2}t\right)+\frac{\mu+\nu}{\mu-\nu}
      \sinh\left(\frac{\mu-\nu}{2}t\right)}, \nonumber \\
F(t)&=&\cosh\left(\frac{\mu-\nu}{2}t\right)+
     \frac{\mu+\nu}{\mu-\nu}\sinh\left(\frac{\mu-\nu}{2}t\right),  
\nonumber \\
G(t)&=&\frac{\frac{2\nu}{\mu-\nu}\sinh\left(\frac{\mu-\nu}{2}t\right)}
     {\cosh\left(\frac{\mu-\nu}{2}t\right)+\frac{\mu+\nu}{\mu-\nu}
      \sinh\left(\frac{\mu-\nu}{2}t\right)}.
\end{eqnarray}
This is a kind of disentangling formula, see for example \cite{KF1} as a 
general introduction.

If from (\ref{eq:disentangling formula}) 
\begin{equation}
\label{eq:Q-D-harmonic oscillator}
\widehat{\tau}(t)
\equiv e^{t(-i\omega_{0}K_{0}+L)}\widehat{\tau}(0)
=e^{\frac{\mu-\nu}{2}t}
e^{G(t)K_{+}}e^{-i\omega_{0}tK_{0}-2\log(F(t))K_{3}}e^{E(t)K_{-}}\widehat{\tau}(0)
\end{equation}
then the original form is given by
\begin{eqnarray}
\label{eq:Q-D-H-O solution}
{\tau}(t)
&=&
\frac{\mbox{e}^{\frac{\mu-\nu}{2}t}}{F(t)}
\sum_{n=0}^{\infty}
\frac{G(t)^{n}}{n!}(a^{\dagger})^{n}
\{
\exp\left(\{-i\omega_{0}t-\log(F(t))\}N\right)\times  \nonumber \\
&&
\left\{
\sum_{m=0}^{\infty}
\frac{E(t)^{m}}{m!}a^{m}\tau(0)(a^{\dagger})^{m}
\right\}
\exp\left(\{i\omega_{0}t-\log(F(t))\}N\right)
\}
a^{n}.
\end{eqnarray}
See \cite{EFS}. Next, we list some results from \cite{FS2} 
for the latter convenience.

\noindent\vspace{3mm}
{\bf (i)}\ If $\tau(0)=\ket{0}\bra{0}$ then
\begin{equation}
\label{eq:example 1}
\tau(t)=\frac{\mbox{e}^{\frac{\mu-\nu}{2}t}}{F(t)}e^{\log G(t) N}.
\end{equation}

\noindent\vspace{3mm}
{\bf (ii)}\ If $\tau(0)=\ket{\alpha}\bra{\alpha}$\ where  
$\ket{\alpha}$ ($\alpha \in \fukuso$) is a coherent state defined by 
\[
\ket{\alpha}
=e^{\alpha a^{\dagger}-\bar{\alpha}a}\ket{0}
=e^{-\frac{|\alpha|^{2}}{2}}e^{\alpha a^{\dagger}}\ket{0}\quad
\left(
\Longleftarrow \ a\ket{\alpha}=\alpha\ket{\alpha}
\right)
\]
then
\begin{equation}
\label{eq:example 2}
\tau(t)
=
\left(1-G(t)\right)
e^{|\alpha|^{2}e^{-(\mu-\nu)t}\log G(t)}
e^{-\log G(t)
\left\{
\alpha e^{-\left(\frac{\mu-\nu}{2}+i\omega_{0}\right)t}a^{\dagger}+
\bar{\alpha}e^{-\left(\frac{\mu-\nu}{2}-i\omega_{0}\right)t}a-N
\right\}
  }.
\end{equation}
The main part (which corresponds to the classical one)
\[
\alpha e^{-\left(\frac{\mu-\nu}{2}+i\omega_{0}\right)t}a^{\dagger}+
\bar{\alpha}e^{-\left(\frac{\mu-\nu}{2}-i\omega_{0}\right)t}a
\]
appears in the formula, see Appendix. 
This derivation is not easy, so see \cite{FS2} for further details.

\vspace{5mm}\noindent
[II]\ Second, we must calculate $e^{tY}$. We decompose $Y$ into two parts
\begin{equation}
\label{eq:decomposition of Y}
Y=-i\Omega (\widetilde{Y}_{1}-\widetilde{Y}_{2})
\end{equation}
where
\begin{eqnarray}
\label{eq:Y-1,Y-2}
\widetilde{Y}_{1}
&=&
\left(
\begin{array}{cccc}
0 & 0 & a\otimes {\bf 1} & 0               \\
0 & 0 & 0 & a\otimes {\bf 1}               \\
a^{\dagger}\otimes {\bf 1} & 0 & 0 & 0  \\
0 & a^{\dagger}\otimes {\bf 1} & 0 & 0
\end{array}
\right)
=\left(
\begin{array}{cc}
                & a \\
a^{\dagger} &  
\end{array}
\right)
\otimes
\left(
\begin{array}{cc}
 {\bf 1} &           \\
          & {\bf 1} 
\end{array}
\right), \nonumber \\
&{}& \\
\widetilde{Y}_{2}
&=&
\left(
\begin{array}{cccc}
0 & {\bf 1}\otimes (a^{\dagger})^{T} & 0 & 0 \\
{\bf 1}\otimes a^{T} & 0 & 0 & 0                 \\
0 & 0 & 0 & {\bf 1}\otimes (a^{\dagger})^{T} \\
0 & 0 & {\bf 1}\otimes a^{T} & 0
\end{array}
\right)
=
\left(
\begin{array}{cc}
 {\bf 1} &          \\
          & {\bf 1} 
\end{array}
\right)
\otimes
\left(
\begin{array}{cc}
                & a \\
a^{\dagger} &  
\end{array}
\right)^{T}.  \nonumber
\end{eqnarray}

\noindent
From (\ref{eq:decomposition of Y}), (\ref{eq:Y-1,Y-2}) and 
\cite{FS} it is easy to see
\begin{eqnarray}
\label{eq:}
e^{tY}
&=&
\mbox{exp}
\left(
-i\Omega t
\left(
\begin{array}{cc}
  0             & a \\
a^{\dagger} & 0
\end{array}
\right)
\right)
\otimes
\mbox{exp}
\left(
i\Omega t
\left(
\begin{array}{cc}
  0             & a \\
a^{\dagger} & 0
\end{array}
\right)
\right)^{T} \nonumber \\
&=&
\left(
\begin{array}{cc}
\cos(\Omega t\sqrt{aa^{\dagger}}) & 
-i\frac{1}{\sqrt{aa^{\dagger}}}\sin(\Omega t\sqrt{aa^{\dagger}})a   \\
-i\frac{1}{\sqrt{a^{\dagger}a}}\sin(\Omega t\sqrt{a^{\dagger}a})a^{\dagger} & 
\cos(\Omega t\sqrt{a^{\dagger}a}) 
\end{array}
\right)
\otimes \nonumber \\
&&
\left(
\begin{array}{cc}
\cos(\Omega t\sqrt{aa^{\dagger}}) & 
i\frac{1}{\sqrt{aa^{\dagger}}}\sin(\Omega t\sqrt{aa^{\dagger}})a   \\
i\frac{1}{\sqrt{a^{\dagger}a}}\sin(\Omega t\sqrt{a^{\dagger}a})a^{\dagger} & 
\cos(\Omega t\sqrt{a^{\dagger}a}) 
\end{array}
\right)^{T}.
\end{eqnarray}

\vspace{5mm}\noindent
[III]\ \ Third, we must calculate $e^{\frac{t^{2}}{2}[X,Y]}$. For the purpose 
we first calculate $[X,Y]$, which is relatively easy. 
Note that $\widetilde{Y}_{1}$ and $\widetilde{Y}_{2}$ commute 
from (\ref{eq:Y-1,Y-2}).

Then the result is
\[
[X,Y]=-i\Omega\left\{[X,\widetilde{Y}_{1}]-[X,\widetilde{Y}_{2}]\right\}
\]
where
\[
[X,\widetilde{Y}_{1}]
=
\left(
\begin{array}{cccc}
  0 & 0 & A & 0  \\
  0 & 0 & 0 & A  \\
  B & 0 & 0 & 0  \\
  0 & B & 0 & 0 
\end{array}
\right),\quad
[X,\widetilde{Y}_{2}]
=
\left(
\begin{array}{cccc}
  0 & C & 0 & 0  \\
  D & 0 & 0 & 0  \\
  0 & 0 & 0 & C  \\
  0 & 0 & D & 0 
\end{array}
\right)
\]
and
\begin{eqnarray*}
A&=& \frac{\mu +\nu }{2}a\otimes {\bf 1} -\nu {\bf 1}\otimes a^{T}, \quad 
B=-\frac{\mu +\nu }{2}a^{\dagger}\otimes {\bf 1}+\mu {\bf 1}\otimes (a^{\dagger})^{T},  \\
C&=&-\nu a^{\dagger}\otimes {\bf 1} +\frac{\mu +\nu }{2}{\bf 1}\otimes (a^{\dagger})^{T}, \quad
D=\mu a\otimes {\bf 1}-\frac{\mu +\nu }{2}{\bf 1}\otimes a^{T}.
\end{eqnarray*}

\noindent
It is easy to see that
\[
[A,C]=[A,D]=0,\quad [B,C]=[B,D]=0,
\]
so we can conclude that $[X,\widetilde{Y}_{1}]$ and $[X,\widetilde{Y}_{2}]$ commute.

Since
\[
e^{\frac{t^{2}}{2}[X,Y]}=
e^{-\frac{it^{2}}{2}\Omega[X,\widetilde{Y}_{1}]}
e^{\frac{it^{2}}{2}\Omega[X,\widetilde{Y}_{2}]}
\]
we can calculate each term easily. The result is
\begin{footnotesize}
\begin{eqnarray*}
&&e^{-\frac{it^{2}}{2}\Omega[X,\widetilde{Y}_{1}]}= \\
&&
\left(
\begin{array}{cccc}
\cos(\frac{t^{2}}{2}\Omega\sqrt{AB}) & 0 & -\frac{i}{\sqrt{AB}}\sin(\frac{t^{2}}{2}\Omega\sqrt{AB})A & 0  \\
0 & \cos(\frac{t^{2}}{2}\Omega\sqrt{AB}) & 0 & -\frac{i}{\sqrt{AB}}\sin(\frac{t^{2}}{2}\Omega\sqrt{AB})A  \\
-\frac{i}{\sqrt{BA}}\sin(\frac{t^{2}}{2}\Omega\sqrt{BA})B & 0 & \cos(\frac{t^{2}}{2}\Omega\sqrt{BA}) & 0  \\
0 & -\frac{i}{\sqrt{BA}}\sin(\frac{t^{2}}{2}\Omega\sqrt{BA})B & 0 & \cos(\frac{t^{2}}{2}\Omega\sqrt{BA})
\end{array}
\right)
\end{eqnarray*}
\end{footnotesize}
and
\begin{footnotesize}
\begin{eqnarray*}
&&e^{\frac{it^{2}}{2}\Omega[X,\widetilde{Y}_{2}]}= \\
&&
\left(
\begin{array}{cccc}
\cos(\frac{t^{2}}{2}\Omega\sqrt{CD}) & \frac{i}{\sqrt{CD}}\sin(\frac{t^{2}}{2}\Omega\sqrt{CD})C & 0 & 0  \\
\frac{i}{\sqrt{DC}}\sin(\frac{t^{2}}{2}\Omega\sqrt{DC})D & \cos(\frac{t^{2}}{2}\Omega\sqrt{DC}) & 0 & 0  \\
0 & 0 & \cos(\frac{t^{2}}{2}\Omega\sqrt{CD}) & \frac{i}{\sqrt{CD}}\sin(\frac{t^{2}}{2}\Omega\sqrt{CD})C  \\
0 & 0 & \frac{i}{\sqrt{DC}}\sin(\frac{t^{2}}{2}\Omega\sqrt{DC})D & \cos(\frac{t^{2}}{2}\Omega\sqrt{DC})
\end{array}
\right).
\end{eqnarray*}
\end{footnotesize}

\vspace{10mm}
In last, we shall restore the result to original form. For the purpose, 
ignoring the term $e^{\frac{t^{2}}{2}[X,Y]}$ we set
\begin{equation}
\label{eq:approximate formula}
\hat{\tilde{\rho}}(t)=e^{tY}e^{tX}\hat{\rho}(0)
=e^{tY}\hat{\tilde{\rho}}_{1}(t), 
\quad 
\hat{\tilde{\rho}}_{1}(t)=e^{tX}\hat{\rho}(0)
\end{equation}
and
\[
\rho(0)=
\left(
\begin{array}{cc}
\rho_{00}(0) & \rho_{01}(0) \\
\rho_{10}(0) & \rho_{11}(0) 
\end{array}
\right).
\]
Then from [I] $\tilde{\rho}_{1}(t)$ becomes
\begin{equation}
\tilde{\rho}_{1}(t)
=
\left(
\begin{array}{cc}
(11) & (12) \\
(21) & (22)
\end{array}
\right)
\end{equation}
where
\begin{eqnarray*}
(11)
&=&
\frac{\mbox{e}^{\frac{\mu-\nu}{2}t}}{F(t)}
\sum_{n=0}^{\infty}
\frac{G(t)^{n}}{n!}(a^{\dagger})^{n}
\{
\exp\left(\{-i\omega_{0}t-\log(F(t))\}N\right)\times  \nonumber \\
&&
\left\{
\sum_{m=0}^{\infty}
\frac{E(t)^{m}}{m!}a^{m}{\bf \rho_{00}(0)}(a^{\dagger})^{m}
\right\}
\exp\left(\{i\omega_{0}t-\log(F(t))\}N\right)
\}
a^{n}, \\
(12)
&=&e^{-i\omega_{0}t}
\frac{\mbox{e}^{\frac{\mu-\nu}{2}t}}{F(t)}
\sum_{n=0}^{\infty}
\frac{G(t)^{n}}{n!}(a^{\dagger})^{n}
\{
\exp\left(\{-i\omega_{0}t-\log(F(t))\}N\right)\times  \nonumber \\
&&
\left\{
\sum_{m=0}^{\infty}
\frac{E(t)^{m}}{m!}a^{m}{\bf \rho_{01}(0)}(a^{\dagger})^{m}
\right\}
\exp\left(\{i\omega_{0}t-\log(F(t))\}N\right)
\}
a^{n}, \\
(21)
&=&e^{i\omega_{0}t}
\frac{\mbox{e}^{\frac{\mu-\nu}{2}t}}{F(t)}
\sum_{n=0}^{\infty}
\frac{G(t)^{n}}{n!}(a^{\dagger})^{n}
\{
\exp\left(\{-i\omega_{0}t-\log(F(t))\}N\right)\times  \nonumber \\
&&
\left\{
\sum_{m=0}^{\infty}
\frac{E(t)^{m}}{m!}a^{m}{\bf \rho_{10}(0)}(a^{\dagger})^{m}
\right\}
\exp\left(\{i\omega_{0}t-\log(F(t))\}N\right)
\}
a^{n}, \\
(22)
&=&
\frac{\mbox{e}^{\frac{\mu-\nu}{2}t}}{F(t)}
\sum_{n=0}^{\infty}
\frac{G(t)^{n}}{n!}(a^{\dagger})^{n}
\{
\exp\left(\{-i\omega_{0}t-\log(F(t))\}N\right)\times  \nonumber \\
&&
\left\{
\sum_{m=0}^{\infty}
\frac{E(t)^{m}}{m!}a^{m}{\bf \rho_{11}(0)}(a^{\dagger})^{m}
\right\}
\exp\left(\{i\omega_{0}t-\log(F(t))\}N\right)
\}
a^{n}
\end{eqnarray*}
and from [II] $\tilde{\rho}(t)$ becomes
\begin{eqnarray*}
\tilde{\rho}(t)
&=&
\left(
\begin{array}{cc}
\cos(\Omega t\sqrt{N+{\bf 1}}) & 
-i\frac{1}{\sqrt{N+{\bf 1}}}\sin(\Omega t\sqrt{N+{\bf 1}})a \\
-i\frac{1}{\sqrt{N}}\sin(\Omega t\sqrt{N})a^{\dagger} & 
\cos(\Omega t\sqrt{N}) 
\end{array}
\right)
\tilde{\rho}_{1}(t)\times \\
&{}&\left(
\begin{array}{cc}
\cos(\Omega t\sqrt{N+{\bf 1}}) & 
i\frac{1}{\sqrt{N+{\bf 1}}}\sin(\Omega t\sqrt{N+{\bf 1}})a \\
i\frac{1}{\sqrt{N}}\sin(\Omega t\sqrt{N})a^{\dagger} & 
\cos(\Omega t\sqrt{N}) 
\end{array}
\right)
\end{eqnarray*}
or by making a slight modification in terms of\  $af(N)=f(N+{\bf 1})a$ 
\begin{eqnarray}
\tilde{\rho}(t)
&=&
\left(
\begin{array}{cc}
\cos(\Omega t\sqrt{N+{\bf 1}}) & 
-i\frac{1}{\sqrt{N+{\bf 1}}}\sin(\Omega t\sqrt{N+{\bf 1}})a \\
-i\frac{1}{\sqrt{N}}\sin(\Omega t\sqrt{N})a^{\dagger} & 
\cos(\Omega t\sqrt{N}) 
\end{array}
\right)
\tilde{\rho}_{1}(t)\times \nonumber \\
&{}&\left(
\begin{array}{cc}
\cos(\Omega t\sqrt{N+{\bf 1}}) & 
ia\frac{1}{\sqrt{N}}\sin(\Omega t\sqrt{N}) \\
ia^{\dagger}\frac{1}{\sqrt{N+{\bf 1}}}\sin(\Omega t\sqrt{N+{\bf 1}}) & 
\cos(\Omega t\sqrt{N}) 
\end{array}
\right).
\end{eqnarray}

By making use of this formula let us calculate an important example. 
The initial state is

\vspace{3mm}\noindent
{\bf Example}
\begin{equation}
\rho(0)=\frac{1}{2}
\left(
\begin{array}{cc}
\ket{0}\bra{0} &                    \\
    & \ket{\alpha}\bra{\alpha}
\end{array}
\right)
\end{equation}
where $\ket{\alpha}$ is a coherent state in [I].

Then the result is
\begin{equation}
\tilde{\rho}_{1}(t)=\frac{1}{2}
\left(
\begin{array}{cc}
A &    \\
   & B
\end{array}
\right)
\end{equation}
where
\begin{eqnarray*}
A&=&\frac{\mbox{e}^{\frac{\mu-\nu}{2}t}}{F(t)}e^{\log G(t) N}, \\
B&=&\left(1-G(t)\right)
e^{|\alpha|^{2}e^{-(\mu-\nu)t}\log G(t)}
e^{
-\log G(t)
\left\{
\alpha e^{-\left(\frac{\mu-\nu}{2}+i\omega_{0}\right)t}a^{\dagger}+
\bar{\alpha}e^{-\left(\frac{\mu-\nu}{2}-i\omega_{0}\right)t}a-N
\right\}
  }
\end{eqnarray*}
(see (\ref{eq:example 1}) and (\ref{eq:example 2})) and
\begin{equation}
\tilde{\rho}(t)=\frac{1}{2}
\left(
\begin{array}{cc}
(11) & (12) \\
(21) & (22)
\end{array}
\right)
\end{equation}
where
\begin{eqnarray*}
(11)
&=&
\cos(\Omega t\sqrt{N+{\bf 1}})A\cos(\Omega t\sqrt{N+{\bf 1}})+
\frac{1}{\sqrt{N+{\bf 1}}}\sin(\Omega t\sqrt{N+{\bf 1}})a Ba^{\dagger}
\frac{1}{\sqrt{N+{\bf 1}}}\sin(\Omega t\sqrt{N+{\bf 1}}), \\
(12)
&=&
i\cos(\Omega t\sqrt{N+{\bf 1}})Aa\frac{1}{\sqrt{N}}\sin(\Omega t\sqrt{N})
-i\frac{1}{\sqrt{N+{\bf 1}}}\sin(\Omega t\sqrt{N+{\bf 1}})a B\cos(\Omega t\sqrt{N}),  \\
(21)
&=&
-i\frac{1}{\sqrt{N}}\sin(\Omega t\sqrt{N})a^{\dagger}A\cos(\Omega t\sqrt{N+{\bf 1}})+
i\cos(\Omega t\sqrt{N})Ba^{\dagger}\frac{1}{\sqrt{N+{\bf 1}}}\sin(\Omega t\sqrt{N+{\bf 1}}), \\
(22)
&=&
\frac{1}{\sqrt{N}}\sin(\Omega t\sqrt{N})a^{\dagger}Aa\frac{1}{\sqrt{N}}\sin(\Omega t\sqrt{N})+
\cos(\Omega t\sqrt{N})B\cos(\Omega t\sqrt{N}).
\end{eqnarray*}

These forms are compact and comparatively beautiful. 
Though we can of course calculate another example we stop here.

\vspace{10mm}
In this paper we reconsidered the Jaynes--Cummings model 
with dissipation from a different point of view and constructed 
a compact approximate solution when some initial condition 
was given. It is very fresh as far as we know. 
We will leave a further construction to readers who are 
interested in this topic. 
As for the preceding works see \cite{KCRS1}, \cite{KCRS2} and 
\cite{KF2}, \cite{KF3}.

We conclude this paper by stating some future prospects. Our real 
target is the following master equation :
\[
\label{eq:Q-D Rabi}
\frac{\partial}{\partial t}\rho=-i[H_{R},\rho]
+
{\mu}
\left\{a\rho a^{\dagger}-\frac{1}{2}(a^{\dagger}a\rho+\rho a^{\dagger}a)\right\}
+
{\nu}
\left\{a^{\dagger}\rho{a}-\frac{1}{2}(aa^{\dagger}\rho+\rho aa^{\dagger})\right\}
\]
where $H_{R}$ is the Rabi Hamiltonian (without RWA (Rotating Wave Approximation)) 
given by
\begin{eqnarray*}
\label{eq:Rabi}
H_{R}
&=&
\frac{\omega_{0}}{2}\sigma_{3}\otimes {\bf 1}+ 
\omega_{0}1_{2}\otimes a^{\dagger}a +
\Omega \sigma_{1}\otimes (a+a^{\dagger})
\\
&=&
\left(
  \begin{array}{cc}
    \frac{\omega_{0}}{2}+\omega_{0}N & \Omega(a+a^{\dagger})   \\
    \Omega(a+a^{\dagger}) & -\frac{\omega_{0}}{2}+\omega_{0}N
  \end{array}
\right).
\end{eqnarray*}

We call this {\bf the Rabi model with dissipation}. 
The Jaynes--Cummings model (which is an approximate model 
with RWA) has some weak points (see for example \cite{JL} and 
its references), so we must treat a more realistic model like this. 
In the following paper(s) we will attack this model.

\vspace{10mm}
\begin{center}
 \begin{Large}
  \textbf{Appendix}
 \end{Large}
\end{center}

\vspace{5mm}
In this appendix we review the solution of classical damped harmonic 
oscillator, which is important to understand the text. 
See any textbook on Mathematical Physics.

The differential equation is given by
\begin{equation}
\label{eq:classical damped harmonic oscillator}
\ddot{x}+\gamma \dot{x}+\omega^{2}x=0\quad (\gamma > 0)
\end{equation}
where $x=x(t),\ \dot{x}=dx/dt$ and the mass is set to 1 for simplicity. 
In the following we treat only the case $\omega > \gamma/2$ (the case 
$\omega=\gamma/2$ may be interesting). 

Solutions (with complex form) are well--known to be
\[
x_{\pm}(t)=e^{-\left(\frac{\gamma}{2}\pm i
\sqrt{\omega^{2}-(\frac{\gamma}{2})^{2}}\right)t},
\]
so the general solution is given by
\begin{eqnarray}
x(t)
&=&
\left\{
\alpha e^{-\left(\frac{\gamma}{2}+i
\sqrt{\omega^{2}-(\frac{\gamma}{2})^{2}}\right)t}
+
\bar{\alpha}e^{-\left(\frac{\gamma}{2}-i
\sqrt{\omega^{2}-(\frac{\gamma}{2})^{2}}\right)t}
\right\}
x(0)  \nonumber \\
&=&
\left\{
\alpha e^{-\left(\frac{\gamma}{2}+i\omega
\sqrt{1-(\frac{\gamma}{2\omega})^{2}}\right)t}
+
\bar{\alpha}e^{-\left(\frac{\gamma}{2}-i\omega
\sqrt{1-(\frac{\gamma}{2\omega})^{2}}\right)t}
\right\}
x(0)
\end{eqnarray}
where $\alpha$ is any complex number.

 If $\gamma/2\omega$ is small enough we have 
 an approximate solution
\begin{equation}
x(t)\approx 
\left\{
\alpha e^{-\left(\frac{\gamma}{2}+i\omega\right)t}
+
\bar{\alpha} e^{-\left(\frac{\gamma}{2}-i\omega\right)t}
\right\}
x(0).
\end{equation}
%



\begin{thebibliography}{99}
%
\bibitem{FS}K. Fujii and T. Suzuki :
\newblock An Approximate Solution of the Jaynes--Cummings 
Model with Dissipation, 
\newblock to appear in Int. J. Geom. Methods Mod. Phys, {\bf 8}, No. 8, 
\newblock arXiv : 1103.0329 [math-ph].
%
%
\bibitem{FHKW1}K. Fujii, K. Higashida, R. Kato and Y. Wada : 
\newblock Cavity QED and Quantum Computation in the Weak Coupling Regime, 
\newblock J. Opt. B : Quantum and Semiclass. Opt, {\bf 6} (2004), 502, 
\newblock quant-ph/0407014. 
%
\bibitem{FHKW2}K. Fujii, K. Higashida, R. Kato and Y. Wada : 
\newblock Cavity QED and Quantum Computation in the Weak Coupling Regime II : 
Complete Construction of the Controlled--Controlled NOT Gate, 
\newblock Trends in Quantum Computing Research, Susan Shannon (Ed.), 
{\bf Chapter 8}, Nova Science Publishers, 2006 and 
Computer Science and Quantum Computing, James E. Stones (Ed.), 
{\bf Chapter 1}, Nova Science Publishers, 2007, 
\newblock quant-ph/0501046. 
%
\bibitem{BP}H. -P. Breuer and F. Petruccione : 
\newblock The theory of open quantum systems, 
\newblock Oxford University Press, New York, 2002.
%
\bibitem{WS}W. P. Schleich : 
\newblock Quantum Optics in Phase Space,
\newblock WILEY--VCH, Berlin, 2001.
%
\bibitem{Scala et al-1}M. Scala, B. Militello, A. Messina, S. Maniscalco, 
J. Piilo and K.-A. Suominen :
\newblock Cavity losses for the dissipative Jaynes-Cummings Hamiltonian 
beyond Rotating Wave Approximation, 
\newblock J. Phys. A: Math. Theor. {\bf 40} (2007), 14527,
\newblock arXiv : 0709.1614 [quant-ph].
%
\bibitem{JC}E. T. Jaynes and F. W. Cummings : 
\newblock Comparison of Quantum and Semiclassical Radiation Theories 
with Applications to the Beam Maser, 
\newblock Proc. IEEE, {\bf 51} (1963), 89.
%
\bibitem{EFS}R. Endo, K. Fujii and T. Suzuki :
\newblock General Solution of the Quantum Damped Harmonic Oscillator, 
\newblock Int. J. Geom. Methods Mod. Phys, {\bf 5} (2008), 653, 
\newblock arXiv : 0710.2724 [quant-ph].
%
\bibitem{FS2}K. Fujii and T. Suzuki :
\newblock General Solution of the Quantum Damped Harmonic Oscillator II :
Some Examples, 
\newblock Int. J. Geom. Methods Mod. Phys, {\bf 6} (2009), 225, 
\newblock arXiv : 0806.2169 [quant-ph].
%
\bibitem{KF1}K. Fujii : 
\newblock Introduction to Coherent States and Quantum Information Theory, 
\newblock quant-ph/0112090.
%
\bibitem{CZ}C. Zachos :
\newblock Crib Notes on Campbell-Baker-Hausdorff expansions, 
\newblock unpublished, 1999,
\newblock see\ http://www.hep.anl.gov/czachos/index.html.
%
\bibitem{KCRS1}A. B. Klimov, S. M. Chumakov, J. C. Retamal and C. Saavedra : 
\newblock An algebraic approach to the Jaynes-Cummings model with dissipation, 
\newblock Phys. Lett. A, {\bf 211} (1996), 143.
%
\bibitem{KCRS2}C. Saavedra, A. B. Klimov, S. M. Chumakov and J. C. Retamal : 
\newblock Dissipation in collective interactions, 
\newblock Phys. Rev. A, {\bf 58} (1998), 4078.
%
\bibitem{KF2}K. Fujii : 
\newblock Algebraic Structure of a Master Equation with Generalized 
Lindblad Form,
\newblock Int. J. Geom. Methods Mod. Phys, {\bf 5} (2008), 1033, 
\newblock arXiv : 0802.3252 [quant-ph].
%
\bibitem{KF3}K. Fujii : 
\newblock A Master Equation with Generalized Lindblad Form and a Unitary 
Transformation by the Squeezing Operator, 
\newblock arXiv : 0803.3105 [quant-ph].
%
\bibitem{JL}Jonas Larson :
\newblock On vacuum induced Berry phases, 
\newblock arXiv:1107.3447 [quant-ph].
\end{thebibliography}
\end{document}